\documentclass[pra,twocolumn,amsmath,amssymb,unsortedaddress,superscriptaddress,showpacs]{revtex4-1}

\usepackage{latexsym,epsfig,graphicx}
\usepackage[caption=false]{subfig}
\usepackage{dcolumn}
\usepackage{bm}
\usepackage{comment}
\pdfoutput=1

\begin{document}

\title{Topological phases, edge modes, and the Hofstadter butterfly in coupled Su-Schrieffer-Heeger systems}

\author{Karmela Padavi$\mathrm{\acute{c}}$}
\thanks{kpadavi2@illinois.edu}
\affiliation{Department of Physics, University of Illinois at
Urbana-Champaign, Urbana, Illinois 61801-3080, USA}
\author{Suraj S. Hegde}
\thanks{shegde2@illinois.edu}
\affiliation{Department of Physics, University of Illinois at
Urbana-Champaign, Urbana, Illinois 61801-3080, USA}
\author{Wade DeGottardi}
\thanks{wdegott@umd.edu}
\affiliation{Joint Quantum Institute, College Park, Maryland 20742, USA}
\author{Smitha Vishveshwara}
\thanks{smivish@illinois.edu}
\affiliation{Department of Physics, University of Illinois at
Urbana-Champaign, Urbana, Illinois 61801-3080, USA}

\begin{abstract}
Motivated by recent experimental realizations of topological edge states in Su-Schrieffer-Heeger (SSH) chains, 
we theoretically study a ladder system whose legs are comprised of two such chains.

We show that the ladder hosts a rich phase diagram and related edge mode 
structure dictated by choice of inter-chain and intra-chain couplings. Namely, we exhibit three distinct 
physical regimes: a topological hosting localized zero energy edge modes, a 
topologically trivial phase having no edge mode structure, and a regime reminiscent 
of a weak topological insulator having unprotected edge modes resembling a 
``twin-SSH'' construction. In the topological phase, the SSH ladder system acts as an analog of the Kitaev chain, which is 
 known to support localized Majorana fermion end modes, with the difference that bound states of the SSH ladder 
 having the same spatial wavefunction profiles correspond to Dirac fermion modes. Further, inhomogeneity in the couplings 
 can have a drastic effect on the topological phase diagram of the ladder system. In particular for quasiperiodic variations 
 of the inter-chain coupling, the phase diagram reproduces Hofstadter's butterfly pattern. We thus identify the SSH ladder system as a 
 potential candidate for experimental observation of such fractal structure.
 \end{abstract}

\maketitle

\section{Introduction}\label{sec:I}
The advent of the Su-Schrieffer-Heeger (SSH) model~\citep{Su72,Heeger88} as a description of organic chains
was a milestone in the study of condensed matter systems in that it offered one of the first realizations of
fractionalization. It was shown that a simple tight-binding chain having two different alternating bond strengths
between lattice sites hosts localized bound states at its ends, enabling fractionalization for charge
at these ends. Subsequently, the model has been studied as a prototype for fractionalization and
associated topological properties characterized by band-structure based invariants and localized edge modes.
Recent cutting edge experimental developments in diverse disciplines have revealed aspects of the SSH model
and related systems in fascinating ways. In cold atomic setting, measurements of topological invariants,
such as the Zak phase have been performed~\citep{Abanin13}. Topologically robust
charge pumping has been observed~\citep{Bloch16, Takahashi16, Lei13} where the transported charge is
quantized and solely determined by the topology of the pump cycle. Equally
striking, topological systems carrying dispersionless edge modes have been
realized in magneto-optical photonic crystals~\citep{Zheng9}, classical acoustic
meta-materials~\citep{Susstrunk47, Peterson17} and even tunable mechanical systems of
granular particles~\citep{Chaunsali17}. Further, the solitonic state distinguishing the topological phase of the SSH model in particular,
has been observed in cold atomic systems~\citep{Meier16, Genske16} through time-of-flight imaging
 and in photonic quantum walks~\citep{Kitagawa12}.

Here, we show that the simplest of next steps in these studies---coupling two SSH chains~\citep{Li14,Zhang17, Zou15}---
gives rise to rich behavior. We investigate salient features of such ``SSH ladder systems",
theoretically studying several physical properties with an eye towards realizing them in the experimental
settings mentioned above. We chart out the phases exhibited by the the SSH ladder and the phase boundaries
separating them as characterized by a gap in the energy spectrum. We determine the topological nature of
these phases. We identify the nature and behavior of bound states localized at the ends of the ladder which
ought to be realizable in cold atomic, photonic and mechanical settings. We focus on three noteworthy
 aspects of the SSH ladder. i) The most general phase
diagram for the SSH ladder hosts phases that go beyond that of a single chain. These phases show
the distinction between completely trivial phases characterized by the absence of edge modes,
robust topological phases having topologically protected edge modes, and an intermediate coupling regime having unprotected
edge modes. ii) The SSH ladder can act as a model of the Kitaev wire by mimicking its traits in the sense of the Kitaev wire hosting Majorana bound
states, which are of much current interest.  iii) The SSH ladder
system has enough degrees of freedom to exhibit marked effects of inhomogeneous couplings; we focus on the case of (quasi)periodicity.

Parallels between the SSH ladder and the Kitaev chain are of particular significance. To elaborate, the Kitaev chain~\citep{Kitaev01} has
come to the forefront as a key model for capturing essential features of topological superconductors.  A distinctive feature of the model is a clean phase
diagram separating trivial phases from topological ones in which the chain harbors end bound states
that are the right combination of particle-like and hole-like excitations for forming  charge neutral Majorana
fermionic states. The phase diagram for the Kitaev chain, the nature of edge modes and their experimental
realization have been thoroughly studied and the Majorana edge states themselves are known to have unique
properties~\citep{Kitaev01, Alicea12, DeGottardi13, Elliott15, Leijnse12}.
For the Kitaev chain, a basis can be chosen so that the system admits a ladder description~\citep{DeGottardi13A}. Remarkably, we find that the
SSH ladder is capable of mimicking this Kitaev chain ladder when subject to restricted couplings. We
demonstrate how these parallels become manifest. A significant distinction is
that the bound states of the SSH ladder are not Majorana fermions but number-conserving states,
either fermionic or bosonic in nature. Nevertheless, the SSH ladder provides a concrete experimentally realizable
system that can map the phase diagram for the Kitaev chain and the spatial structure of its localized bound edge states.

Extensive work on the Kitaev chain has revealed the manner in which inhomogeneities can greatly alter topological phase diagrams, for instance, periodic variations or disorder~\cite{DeGottardi13A}.
Borrowing from these insights, we show that indeed spatially varying coupling in the SSH ladder system can produce dramatically
different phase diagrams. We employ a transfer matrix technique to directly target the fate of localized end modes in the
presence of such spatial variations. Further, we find that quasiperiodic variations across the ladder reflect the mathematical
structure of Harper's equation and its physical manifestation in Hofstadter's butterfly pattern~\citep{Hofstadter76} --
one of the most well-known examples of fractals in condensed matter physics. While cold atomic systems are well
suited for clean achievement of such self-similar patterns (otherwise experimentally rather challenging in electronic systems), experiments concerning
the Hofstadter Hamiltonian so far~\citep{Aidelsburger13,Miyake13,Osterloh05, Jaksch03} have not included any direct measurements
of its fractal nature. As the topological phases of the SSH ladder can be identified through direct observation of edge wavefunction
spatial profiles, our proposal provides a novel possibility of probing this fractal diagram through time-of-flight imaging.

In what follows, in Sec.~\ref{sec:modelandmethods} 
we introduce the SSH ladder Hamiltonian and outline the use of the momentum 
space dispersion relation, the chiral index topological invariant and the 
tranfer matrix formalism in obtaining its phase diagram. In Sec.~\ref{sec:decoupled} 
we briefly review the properties of a single, uncoupled, SSH chain and proceed 
to discuss the more complex phase diagram of the coupled system in 
Sec.~\ref{sec:phases}. In this section, we identify three distinct regimes (topological, topologically trivial and weakly topological) as 
determined by differing edge mode structures. In Sec.~\ref{sec:kitaev} we focus 
on the phase of the ladder hosting edge modes with wavefunctions having spatial 
profiles matching those of the localized Majorana modes of the Kitaev chain. 
Further, in Sec~\ref{sec:periodic} we employ transfer matrix methods to discuss 
the response of this Kitaev chain analog phase to inhomogenous couplings and 
disorder. Finally, in Sec.~\ref{sec:finitesize} we bring our discussion closer 
to experimental studies of SSH models by considering finite size effects and 
offer an outlook on possible future experimental work in Sec.~\ref{sec:outlook}.

\section{Model and Methods}
\label{sec:modelandmethods}

\begin{figure}[htbp]
\centering
\includegraphics[width=0.5\textwidth]{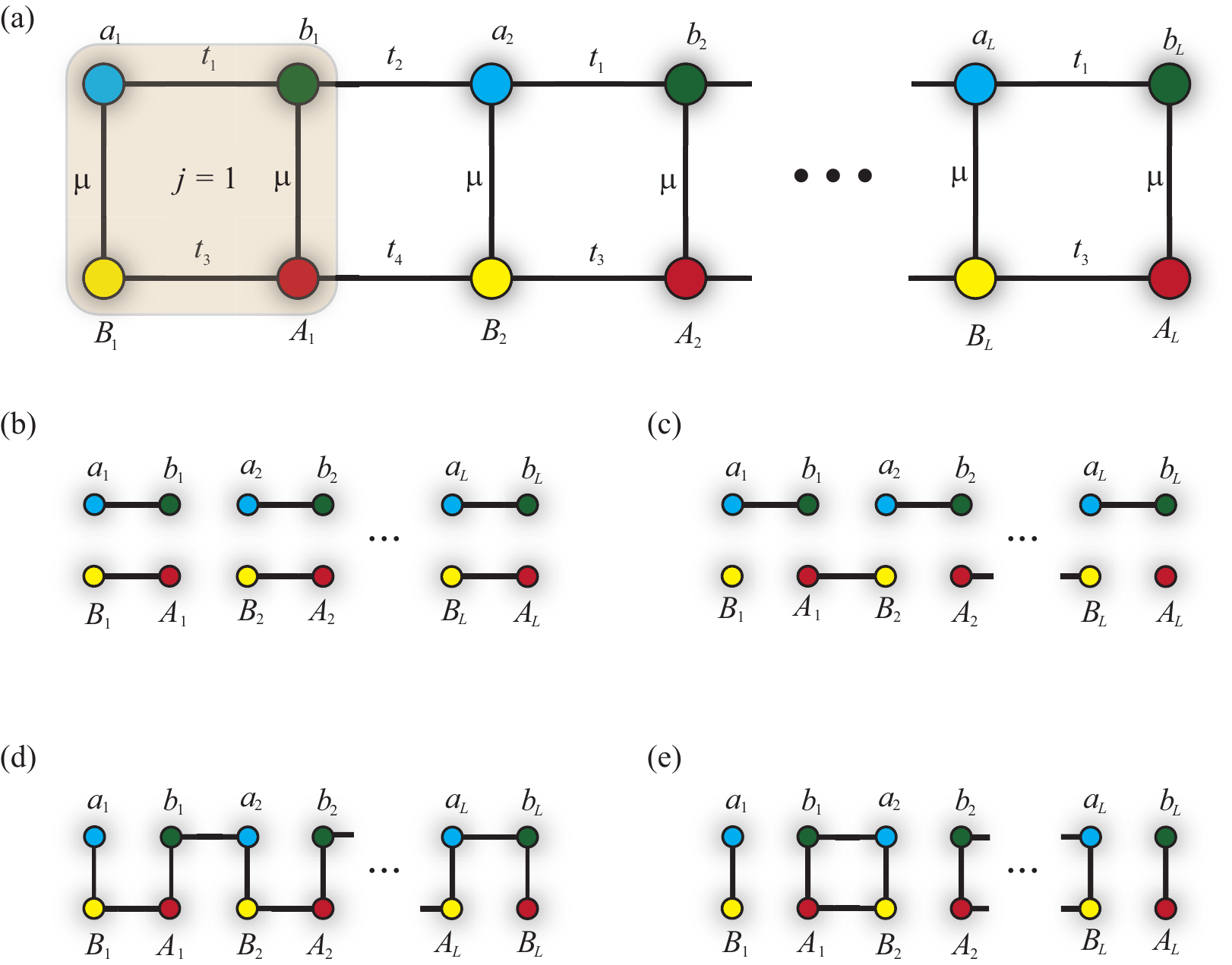}
\caption{Pictorial representation of the fermion couplings in the SSH ladder model, as
represented by the Hamiltonian (1) for (a) generic couplings, (b) for the
decoupled case with the special values of couplings $t_2 = t_3 = 0$ characterized by
$N_S = 1$ exhibiting a single fermion end mode, (c) coupled SSH ladder for
$t_1 = t_4 = 0$ and $t_2 = t_3 = t$. For $|\mu | < 2 | t |$, the
system exhibits a Kitaev-like topologically protected zero mode.
(e) For $t_1 = t_3 = 0$ and $|\mu| \ll t_2,t_4$, the system exhibits mid-gap
states that are not topologically protected.}
\label{fig:ladder}
\end{figure}

The system of interest, shown in Fig.~(\ref{fig:ladder}), is a fermionic ladder composed of two coupled SSH chains.
It exhibits a particle-hole symmetry which arises from the bipartite nature of the Hamiltonian. Consequently,
 its energy spectrum is symmetric about zero energy. As detailed in Sec. \ref{sec:kitaev}, when
the couplings respect certain constraints, the system is a close analog to the celebrated superconducting Kitaev chain;
particle-hole symmetry in this case is a natural consequence of superconductivity, as captured by the Bogoliubov-de Gennes form of its Hamiltonian.
The close relationship  between the two systems gives rise to a clear correspondence between their respective topological phases.
As will be shown, however, rather than hosting Majorana fermions, the topological phases of the SSH ladder
are characterized by Dirac fermions localized at the system edge. As detailed in Sec. \ref{sec:phases}, the more generalized
ladder system shows a rich phase diagram. We note here that several of single-particle properties studied in what follows apply
to bosonic systems as well. Additionally, similar models have been explored in Ref. ~\citep{Li14,Zhang17, Zou15}.

\emph{{\bf Model---}} The unit cell for the SSH ladder is a plaquette  composed of four fermions: for the $j^{\textrm{th}}$ plaquette,
the operators $a_j$, $b_j$ are fermion annihilation operators for sites on the top leg, while the operators $B_j$
and $A_j$ reside on the bottom leg. The Hamiltonian is given by
\begin{eqnarray}
&& H=\sum_{j}(t_1b_{j}^{\dagger}a_{j}+t_2a^{\dagger}_{j+1}b_{j}+t_3A^{\dagger}_{j}B_{j} + t_4
B^{\dagger}_{j+1}A_{j}\nonumber\\
&& + {\mathrm{h.c.}}) + \mu\sum_{j}(a^{\dagger}_{j}B_{j}+
 b^{\dagger}_{j}A_{j}+{\mathrm{h.c.}}).
 \label{eq:genHam}
\end{eqnarray}
The Hamiltonian $H$ is characterized by five real couplings: four hopping amplitudes $t_1$, $t_2$, $t_3$, and $t_4$ and an
inter-chain coupling $\mu$. The system is bi-partite in that it decouples into two interpenetrating
sub-lattices, $a/A$ and $b/B$ coupled to each other but not within themselves. This bi-partite
nature of the model guarantees that the spectrum respects particle-hole symmetry. Fourier transforming Eq.~(\ref{eq:genHam}) yields
\begin{eqnarray}\label{eq:HamBlocks}
&& H_k=\frac{1}{2}\tau_z\otimes[((t_1-t_3)+(t_2-t_4)\cos k)\sigma_x\nonumber\\
&& + (t_2+t_4)\sin k \, \sigma_y] +\frac{1}{2}\mathbb{I}\otimes[((t_1+t_3)+(t_2+t_4)\cos k)\sigma_x \nonumber\\
&& + (t_2-t_4)\sin k \, \sigma_y]+\mu\tau_x\otimes\sigma_x
\end{eqnarray}
where the Pauli matrices $\tau_i$ and $\sigma_i$ act in the ladder leg and sublattice spaces, respectively.

Our analyses of the phases and their properties primarily employ three methods. i) For systems
respecting translational invariance, energy dispersions obtained in momentum space identify
gapless energy contours which delineate phase boundaries. ii) A topological invariant enables
us to identify topological aspects of the phases. iii) A transfer matrix technique charts out
the existence and nature of possible localized end bound states in some phases.
We proceed to elaborate on each of these methods.

\emph{{\bf Dispersion---}} First, to establish phase boundaries, we consider
the dispersion relation corresponding to Eq.~(\ref{eq:HamBlocks}).
Contours in parameter space along which the bulk energy gap closes delimit different phases of the system.
In the most general case, the SSH ladder has a four-band energy dispersion
\begin{eqnarray}\label{eq:GenHEvals}
 && E^2=\mu^2+\frac{1}{2}(|\rho_1|^2+|\rho_2|^2)\nonumber\\
 && \pm\frac{1}{2}\sqrt{(2\mu^2+|\rho_1|^2+|\rho_2|^2)^2+4(\rho_1\rho_2-\mu^2)(\mu^2-\rho^{*}_1\rho^{*}_2)}\nonumber\\
\end{eqnarray}
where
\begin{eqnarray}\label{eq:rho12}
  && \rho_1=\rho_1(k)=t_1+t_2e^{-ik}\nonumber\\
  && \rho_2=\rho_2(k)=t_3+t_4e^{+ik}.
\end{eqnarray}
Here, the gap in the energy spectrum closes at $k=0$ for $\mu^2=(t_1+t_2)(t_3+t_4)$ while the same happens for $\mu^2=(t_1-t_2)(t_3-t_4)$
and $k=\pi$. Additionally, the energy spectrum is gapless for $t_1t_4=t_2t_3$ and $k=(1/2)\cos^{-1}(\mu^2-t_1t_3-t_2 t_4)/(t_1t_4+t_2t_3))$.

Following Ref.~\citep{Wakatsuki14}, we parameterize the couplings in Eq.~(\ref{eq:genHam})
\begin{eqnarray}
&& t_1=(t+\Delta)(1-\eta),\phantom{a}t_2=(t-\Delta)(1+\eta), \nonumber\\
&& t_3=(t-\Delta)(1-\eta), \phantom{a}t_4=(t+\Delta)(1+\eta).
 \label{eq:ParaW}
\end{eqnarray}
with $|\eta| < 1$, $|\Delta| < t$. In the next section, we offer a physical interpretation for these
two parameters in terms of competing energy gaps. Various
topological and topologically trivial phases of the ladder are separated by regions in the $(\mu, \Delta, \eta, t)$ parameter space satisfying
\begin{eqnarray}
 && \mu^2=4(t^2-\Delta^2\eta^2), \nonumber\\
 && \mu^2=4(t^2\eta^2-\Delta^2),\ \mathrm{for} \ t|\eta|>|\Delta|\nonumber\\
 && \Delta=0.
\label{eq:genPB}
\end{eqnarray}

\emph{{\bf Topological invariant---}} Phases of the SSH ladder can be characterized by a
topological invariant. The over-arching Hamiltonian of Eq.~(\ref{eq:genHam}) belongs to
the BDI symmetry class and has phases described by a $\mathbb{Z}-$valued topological
invariant. 
Following Refs.~\citep{Gurarie11, Wakatsuki14} we consider the chiral index
associated with a generic symmetry operator $S$ with the properties $SH_kS^{-1}=-H_k$, $S^2=\mathbb{I}$. The
topological invariant is given by
\begin{eqnarray}\label{eq:topN}
N_S=\mathrm{Tr}\int_{-\pi}^{\pi}\frac{dk}{4 \pi i}S g^{-1}\partial_k g
\end{eqnarray}
where $g(k)=H^{-1}(k)$ is the Green's function at zero energy.

\emph{{\bf Transfer matrix---}} The transfer matrix formalism extensively used in studies of localized states in 1D systems, especially in the presence of
potential landscapes, is applicable to this ladder model. We employ this formalism to study the effects of spatial
modulation of the inter-chain coupling $\mu$ in Sec.~\ref{sec:periodic}. This
method has been further developed as a tool in studies of Majorana modes in the Kitaev wire by two of the authors in
Refs.~\citep{DeGottardi13, DeGottardi13A,DeGottardi11}. 

Within this formalism, the presence of
localized zero energy edge modes is determined by the growth or decay of the eigenfunctions of the transfer matrix.
To construct the matrix, we start with the zero-energy Heisenberg equations of motion, $[a_j,H]=0$, $[b_j,H]=0$, for
\begin{eqnarray}\label{eq:KitHam}
 && H=-\sum_{j}[t_1(b^{\dagger}_{j+1}a_{j}+B^{\dagger}_{j+1}A_{j})+t_2(a^{\dagger}_{j+1}b_{j}+A^{\dagger}_{j+1}B_{2,j})]\nonumber\\
&&  +\sum_{j}\mu_j a^{\dagger}_{j}B_{j}+\mathrm{h.c.}\nonumber
\end{eqnarray}
where the inter-chain coupling $\mu_j$ now varies throughout the ladder. The zero
energy equations of motion couple the $a_i$ and $A_i$ fermions and the $b_i$ and $B_i$ fermions, but
these two sets of equations have no mutual couplings. 
The equations of motion for operators $a_j$ are second-order difference equations of the form
 \begin{equation}
t(1+\delta) a_{j-1} + t(1 -\delta)a_{j+1} + \mu_j A_{j} =0.
\label{weqa}
\end{equation}
 Here we have taken $t_1=t(1-\delta)$ and $t_2=t(1+ \delta)$
This can be cast into the form of transfer matrices:
\begin{equation}\label{eq:tmatA}
   \left(\begin{array}{c}
  a_{j-1}\\
  A_{j}\end{array}\right)=A_{j}\left(\begin{array}{c}
  A_{j}\\
  a_{j+1}\end{array}\right), \
  \mathcal{A}_j=\left(\begin{array}{cc}
  \frac{\mu_n}{t(1+\delta)} &
  -\left(\frac{1-\delta}{1+\delta}\right)\\
  1&0\end{array}\right).
\end{equation}
Thus, the transfer matrix relates wavefunction amplitudes between one slice of the ladder and its adjacent one.
Its multiplicative nature allows us to study the manner in which the wavefunction varies along the ladder.
For localized states, such variation is an exponential decay and can be ascertained based on the eigenvalue
structure of the transfer matrix. Such an analysis will be presented in more detail and used extensively in following sections involving
inhomogenous variations in the SSH ladder (Sec.~\ref{sec:periodic}).

\section{Single SSH chain}
\label{sec:decoupled}

Before discussing the intricacies of the full model, we briefly recapitulate the properties of a
single SSH chain having coupling $t_1$ and $t_2$. The chain possesses two topologically distinct phases where
the topological and trivial phases are characterized by the presence and absence of end zero modes,
respectively. For instance, in the case in which
$t_2 = 0$, and $t_1 \neq 0$, the system is described by pairs of coupled fermions. In contrast,
for $t_1 = 0$ and $t_2 \neq 0$, the system forms pairs of dimers with the exception of the
modes $a_1$ and $b_L$ which are left uncoupled to the rest of the chain. Particle-hole symmetry
guarantees that they are zero energy modes.

The existence of these end bound state fermions is protected by the existence of the bulk energy gap. Tuning away from the special case in
which $t_1 = 0$, the end zero mode is a linear combination of modes near the edge of the system. These modes
 persist as long as the bulk gap does not close. The spectrum of the single SSH chain is given by
$E =\pm\sqrt{t_1^2+t_2^2+2t_1t_2\cos k}$, and thus the bulk gap closes at $k = \pm \pi$ for $t_1 = t_2$. This condition separates the trivial phase from the topological phases. For $t_1 > t_2 > 0$, the system is trivial. For $t_2 > t_1 > 0$,
the system is topological and possesses a single Dirac fermion at each of its ends.

We note that our assignment of phases as topological and topologically trivial
depends on where, within the unit cell, we propose a measurement to detect the zero edge state
should occur (where a ``cut'' occurs). In other words, determining the wavefunction spatial profile at a
site off-set by half of the unit cell width would result in a different
characterization of the phases of the SSH ladder. However, the two previously
discussed phases, remain \emph{topologically distinct} regardless. Below, we use
a convention consistent with studying the zero mode wavefunctions at the
left-most and right-most sites of the SSH ladder.

\begin{figure}[]
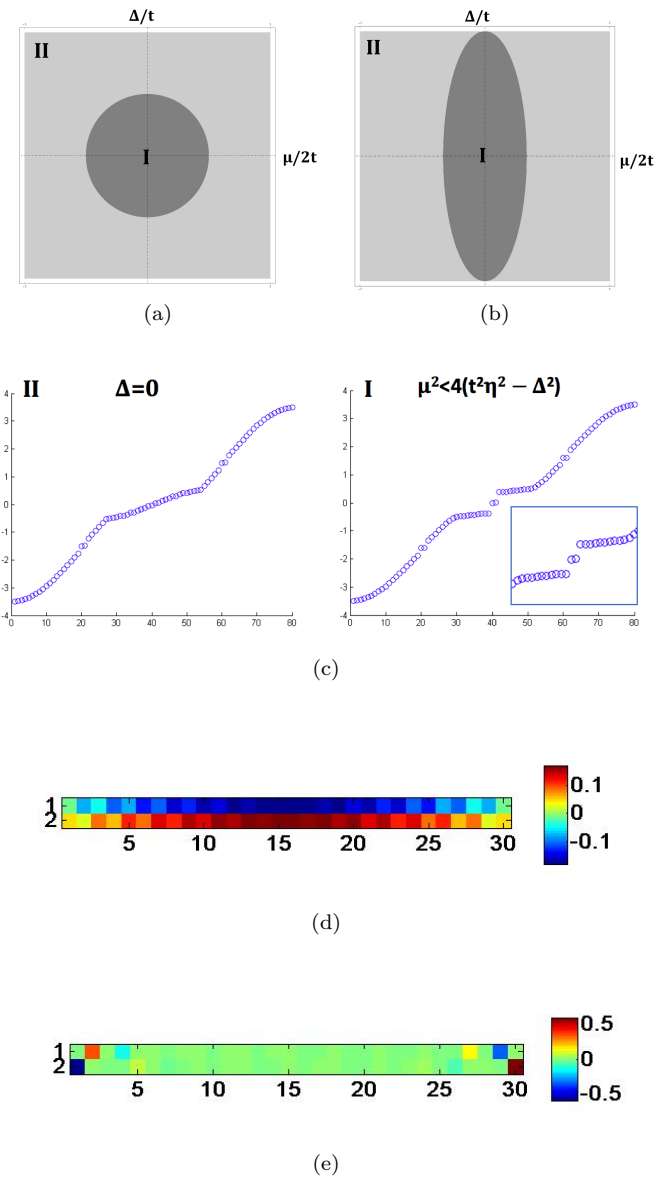

\subfloat[][]{\includegraphics[width=0.25\textwidth]{CircleDiagram2.pdf}}
\subfloat[][]{\includegraphics[width=0.25\textwidth]{EllipseDiagram.pdf}}\\
\subfloat[][]{\includegraphics[width=0.5\textwidth]{Fig2c.pdf}}\\
\subfloat[][]{\includegraphics[width=0.5\textwidth]{Fig2d.pdf}}\\
\subfloat[][]{\includegraphics[width=0.5\textwidth]{Fig2e.pdf}}
\caption {(a) Phase diagram of the general SSH ladder for $|\Delta|<|\eta|t$. This parameter regime corresponds to a ladder configuration composed
of two \emph{identical} SSH chains, as in Fig.~\ref{fig:ladder} (b). Here, phase I hosts a single pair of zero energy edge modes while phase II
is topologically trivial having no zero edge mode structure. (b)  Phase diagram of the general SSH ladder for $|\Delta|>|\eta|t$. Here, the SSH ladder
is in a regime illustrated by Fig.~\ref{fig:ladder} (c) where the two SSH chains that comprise it are \emph{offset} with respect to the relative strength of couplings:
if the bottom chain starts with a weaker coupling then the top starts with the stronger one. As in (a), phase I
hosts a single pair of zero energy edge modes (shown in the inset) and phase II supports no
zero edge modes. (c) Spatial wavefunction profiles and energy spectra,
obtained by numerical diagonalization of the Hamiltonian Eq.~(\ref{eq:HamBlocks}), corresponding to the phases in (a) and (b). In either parameter regime
$\Delta=0$ describes a gapless line in the phase diagram, having the same dispersion relation as the topologically trivial
phase II. The sign of $\Delta$ determines whether $a_i, A_i$ or $b_i, B_i$ modes are localized to a particular
side of the ladder. }
\label{fig:Model34}
\end{figure}

\begin{figure}[]
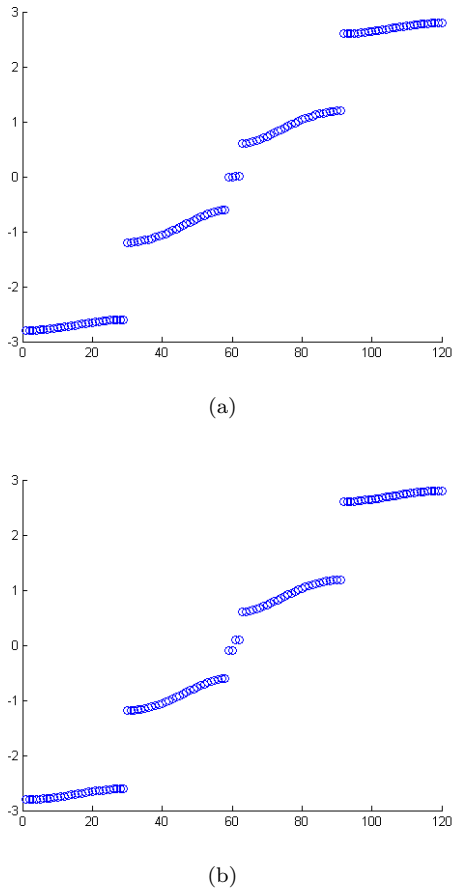

\centering
\subfloat[][]{\includegraphics[width=0.4\textwidth]{Mu0.pdf}} \\
\subfloat[][]{\includegraphics[width=0.4\textwidth]{MuN0.pdf}}
\caption {Energy spectra obtained by numerical diagonalization of the Hamiltonian Eq.~(\ref{eq:HamBlocks}), for (a) a fully
decoupled SSH ladder $\mu=0$ and (b) a weakly coupled $\mu \ll 1$ SSH ladder. A non-zero value of the inter-chain coupling $\mu$
causes the four zero energy states to hybridize, resulting in non-zero energy mid-gap edge modes.}
 \label{fig:4To2}
\end{figure}

\section{Phases of the SSH Ladder}
\label{sec:phases}

As suggested by the phases of the SSH chain, the topology of the ladder
system depends sensitively on the relative strengths of the couplings $t_1, t_2, t_3,$ and $t_4$.
As discussed below, we find three distinct physical regimes: a topological phase analogous to the Kitaev 
chain hosting localized zero energy edge modes, a topologically trivial phase and a regime analogous to a 
weak topological insulator having unprotected edge modes resembling a ``twin-SSH'' construction.
To provide an intuitive picture for these phases, in considering Fig.~\ref{fig:ladder}, it is clear that 
in addition to a topologically trivial phase having no localized edge modes
(Fig.~\ref{fig:ladder} (b)) the SSH ladder can host a single localized mode at 
each edge (Fig.~\ref{fig:ladder} (c)) or two pairs of localized modes at each edge as in Fig.~\ref{fig:ladder} 
(e).

The parameters $\Delta$ and $\eta$ introduced in Eq.~(\ref{eq:ParaW}) quantify the relative
strengths of the various dimerization patterns that the SSH can exhibit. The quantity $\eta$
is indicative of the relative strength of inter-plaquette and intra-plaquette couplings, while $\Delta$ is
a measure of how strongly the upper and lower legs of the ladder have opposite dimerization
patterns. In particular, increasing $\Delta$, for $t, \Delta > 0$ tends to give rise to a single
end mode on each end of the chain. This mode has weight on the $a_i, A_i$ modes on the
left-hand side of the chain and the $b_i, B_i$ modes on the right-hand side. Similarly,
decreasing $\Delta$ for $\Delta < 0$ tends to localize a single fermion with the roles
of $a_i, A_i$ and $b_i, B_i$ interchanged. The parameter $\eta$ treats the upper and
lower legs symmetrically. For $\eta > 0$, there is a tendency to localize two fermionic modes to each end of the ladder.

The topological nature of the phases can be characterized by tailoring the topological invariant introduced in Eq.~\ref{eq:topN} to the SSH ladder.
In particular, identifying $S=\mathbb{I}\otimes\sigma_z$, we obtain
\begin{eqnarray}\label{eq:NVspec}
&& N_S =-\int_{-\pi}^{\pi}\frac{dk}{2 \pi i}\partial_k\log(t_1t_3+t_2t_4+t_2t_3e^{-ik}\nonumber\\
&& + t_1t_4e^{ik}-\mu^2).
\end{eqnarray}
The topological index $N_S$ is thus the winding
number of an ellipse in the complex plane. If the ellipse does not enclose the origin, the system is trivial and $N_S = 0$. This occurs for
\begin{eqnarray}
\mu^2 > (t_1+t_2)(t_3+t_4)=4(t^2-\Delta^2\eta^2).
\label{eq:toptrivcond}
\end{eqnarray}
For
\begin{eqnarray}
&& (t_1-t_2)(t_3- t_4)=4(t^2\eta^2-\Delta^2) < \mu^2 \\
&& < (t_1+t_2)(t_3+t_4)=4(t^2-\Delta^2\eta^2),
\end{eqnarray}
the ellipse encloses the origin and $N_S = \mbox{sgn}(t \Delta)$. The sign of $N_S$
thus determines whether $a_i, A_i$ or $b_i, B_i$ modes are localized to a particular
side of the ladder. The two possible cases are therefore topologically distinct and
separated by a gapless line $\Delta=0$ in the topological phase diagram as in, for instance, Fig.~(\ref{fig:Model34}).

Given that the ladder is essentially two coupled SSH chains, \emph{a priori} the system
seems capable of hosting zero, one or two localized modes at its ends. Our
analysis of the topological invariant $N_S$ shows that this expectation is not quite accurate.
We find that there are two distinct topological phases which exhibit single fermionic zero modes at
each end of the ladder. At the same time, the ladder system cannot host more than one fermionic
mode at each end. In particular, the form of $N_S$ in Eq.~(\ref{eq:NVspec}) makes it clear
that only if the ladder has longer range hopping can $|N_S| > 1$.

We can also address the fate of the zero modes for two topological SSH chains which
are coupled. The two end modes can hybridize, thus forming the analog of a weak
topological insulator, as shown in~\ref{fig:ladder} (e). A single fermion mode is however
protected given that for a spectrum with particle hole symmetry and an odd number of
states there must be at least one zero energy state. This argument is readily
generalized: for a system of $N$ chains coupled by weak inter-channel couplings, we
anticipate that a topological phase (with a single topological edge mode) will result if
and only if the number of chains that are topological is odd, a natural consequence of fermion parity and particle-hole symmetry.

We remark that while the hybridization of edge states in the case of two coupled chains does not allow for a topological
phase hosting zero energy edge states, the resulting regime is distinct from the trivial
regime in that there is a persistent, non-zero energy, bound state at each ladder end. As shown in Fig.~\ref{fig:4To2},
hybridization causes the bound edge states to move away from zero energy but they still form low-energy
mid-gap states. Transitioning into the trivial state having no boundary modes whatsoever requires the closing of the bulk energy gap.

\begin{figure}[]
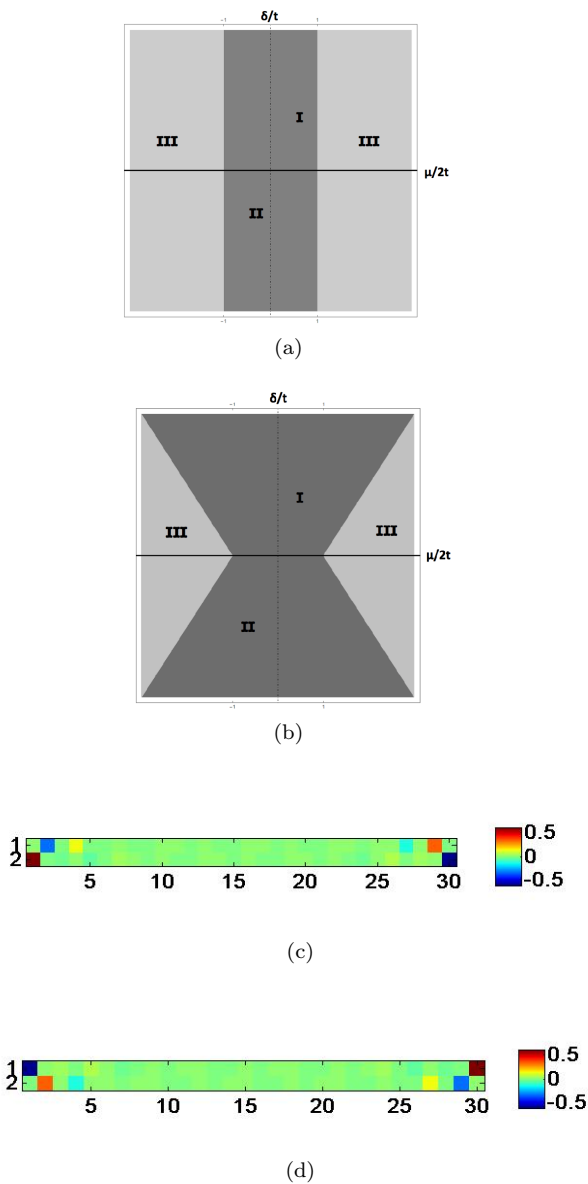

\centering
\subfloat[][]{\includegraphics[width=0.25\textwidth]{T1T2.pdf}}\\
\subfloat[][]{\includegraphics[width=0.25\textwidth]{FiniteSize.pdf}}\\
\subfloat[][]{\includegraphics[width=0.5\textwidth]{Fig4c.pdf}}\\
\subfloat[][]{\includegraphics[width=0.5\textwidth]{Fig4d.pdf}}
\caption{(a) Phase diagram for an infinitely long SSH ladder with $t_1=t_3$ and $t_2=t_4$, (B) phase diagram for
an SSH ladder with hopping parameters as in (a) and a finite size $L$. The slope of the phase boundaries is linear in $L$. (c) and (d) Spatial wavefunction profiles
corresponding to the two topologically non-trivial phases (I and II in (a)) of the reduced SSH ladder.}
 \label{fig:Model12}
\end{figure}

\section{Kitaev Chain Analog}
\label{sec:kitaev}

We now specialize to the case in which $\eta = 0$. We take $t_{1}=t_{4}=t(1-\delta)$,  $t_{2}=t_{3}=t(1+\delta)$, where $\delta = \Delta/t$. As
discussed in the previous section, $\delta$ measures the tendency of the SSH ladder to have the opposite dimerization patterns on the upper
and lower legs, respectively. Thus, larger $|\delta|$ tends to stabilize topologically non-trivial phases.

This special, restricted set of couplings offers a means of probing the phase diagram of the Kitaev chain, a prototype for
studying Majorana wires and the exciting physics of Majorana fermionic bound states~\cite{Kitaev01, Hegde16}. The
Kitaev chain consists of a single chain of electrons possessing on-site local chemical potential $\mu$ that can tunnel
between sites (for instance, with strength $t$) and experience nearest neighbour $p$-wave pairing (for instance, of strength $\Delta$). When
represented in terms of pairs of Majorana fermions on each site, the physical construction exactly maps on to the SSH ladder
with the identification $t_{1/2} = t \pm \Delta$~\cite{DeGottardi13A}. Electronic
bound states at the ends of the SSH ladder map to Majorana fermionic bound states at the end of the Majorana wire.
Most importantly, the dispersions for both systems are identical and the phase boundaries serve to demarcate topological phases from trivial ones.

Specifically, in the Kitaev chain analog limit, the energy dispersion of the SSH ladder takes the form
\begin{equation}\label{eq:1Deval}
E_{\pm}(k)=\pm t \sqrt{(2\cos(k)+\mu/t)^2+(2\delta \sin(k))^2}.
\end{equation}
As in Sec.~\ref{sec:phases}, the loci of energy gap closing points form the phase boundaries between different
topological phases of the system and are given by : $\mu= \pm 2t$ at $k=0,\pi$ and $\delta=0$ at
$k=\cos^{-1}(-\mu/2t)$. The phase diagram determined by these curves is shown in Fig.~\ref{fig:Model12} (a). The topology
of the various parts of the phase diagram are readily established by considering simple cases. The $\delta = - 1$ corresponds
to the case shown in Fig.~\ref{fig:ladder} (d) which can be viewed as a single SSH chain weaving between the upper and lower legs of the ladder.
In accordance with findings for a single, disconnected, chain the system is topologically non-trivial in this case provided that $|\mu| < 2 |t|$.

To re-emphasize the connections and differences between the SSH ladder and the Kitaev chain, the Dirac fermionic end
modes in the SSH ladder correspond to phases with end Majorana modes in the Kitaev chain. 
Though sometimes attributed to the $\mathbb{Z}_2$
character of the Kitaev chain (a system in class D), we see
that this property is actually due to particle-hole symmetry and fermionic parity. Indeed, when the
superconducting order parameter is taken to be real in the Kitaev chain, the Hamiltonian is also in the BDI class.

\section{Inhomogeneous Couplings and the Hofstadter Butterfly}
\label{sec:periodic}

We now consider the effect of including spatial inhomogeneity in a specific coupling terms in the SSH ladder system.
In particular, we remain within the Kitaev chain analog limit of $t_1=t_4$, $t_2=t_3$ and  consider periodicity, quasi-periodicity, and disorder in the inter-chain coupling $\mu$. Our reason
for this is three-fold. First, we wish to preserve particle-hole symmetry and the bipartite nature of the lattice and
thus do not include an actual on-site chemical potential. Second, variation in $\mu$ for the ladder system exactly
maps on to such a variation in a potential landscape in the Kitaev chain, thus enabling us to
parallel Majorana fermion physics in the presence of potential landscapes and disorder~\cite{Sau13, DeGottardi13A, Rieder13, Rieder14, Neven13}.
Third, of the vast parameter space for inhomogeneities, we narrow our study to the most natural choice and show
the rich phase diagram structure stemming from even varying a single parameter.

Explicitly, the inter-chain coupling in our model Hamiltonian of Eq. (\ref{eq:genHam}) takes the form
\begin{eqnarray}
\sum_{j = 1} (\mu_{2j-1} a^{\dagger}_j B_j + \mu_{2j} b^{\dagger}_j A_j + \mathrm{h.c.}).
\end{eqnarray}

The topology of the disordered chain is most conveniently found by employing the transfer matrix method, discussed in Sec. \ref{sec:modelandmethods}. We
define the Lyapunov exponent of the transfer matrix is $\gamma(\{\mu_j\},\delta,L)= \lim_{L \rightarrow \infty}1/L \ln|\lambda|$,
where $\lambda$ is its highest eigenvalue. The exponent is the
inverse of the localization length of the edge-mode wavefunction in the topological phase. The phase
boundaries separating the topologically trivial and non-trivial phases are thus determined by $\gamma=0$, 
corresponding to a diverging localization length at the transition point. 

To probe the fate of the topological phase diagram when inhomogeneities are present, 
we make use of a similarity transformation as given in Ref.~\citep{DeGottardi13}. Explicitly, the similarity
transformation on the full chain transfer matrix is given by
  \begin{eqnarray}
  \mathcal{A}(\mu_n,\delta)= \bigg(\frac{1-\delta}{1+\delta} \bigg)^{N/2} S
  \tilde{\mathcal{A}}(\mu_n/\sqrt{1-\delta^2},\delta=0)S^{-1} \phantom{word}
  \label{simtrans}
\end{eqnarray}
where $S= \rm{diag}( \ell_{\delta}^{1/4},1/\ell_{\delta}^{1/4})$ and $\ell_{\delta} = \frac{1-\delta}{1+\delta}$ and
we have set $t=1$. The matrix $\tilde{\mathcal{A}_n}$ is  the
transfer matrix for a normal tight-binding model -- in the context of quasi-periodicity, a Harper model -- in the absence of a
dimerization $\delta$. The model's on-site chemical potential terms are re-scaled by the
transformation $\mu_n \rightarrow \mu_n / \sqrt{1-\delta^2}$. This map allows the Lyapunov exponent to be
written as a sum of two components, $\gamma(\mu, \delta)= \gamma_{\delta} + \gamma_0$,
one that depends purely on the dimerization $\gamma_{\delta}$ and $\gamma_0 (\mu/\sqrt{1-\delta^2})$, which is the Lyapunov exponent for a system having no dimerization and a rescaled locally varying chemical potential. Thus, knowing wavefunction properties for non-dimerized system enables identifying localized wavefunctions in an SSH ladder having the same spatial variation.

\begin{figure}
\begin{center}
\includegraphics[width=0.45\textwidth]{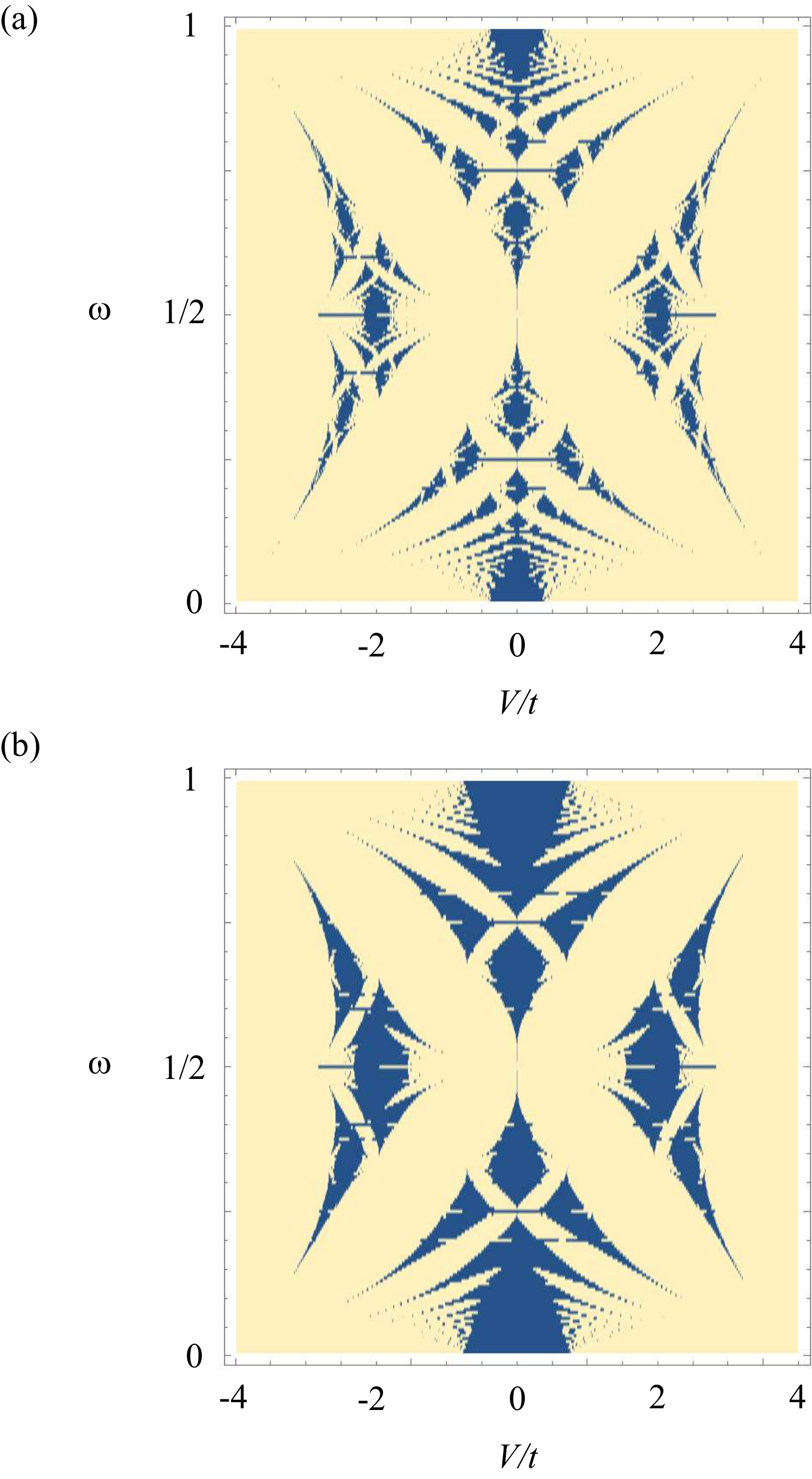}
\caption{The phase diagram for a quasiperiodically varying inter-chain coupling $\mu_n/t = V/t + 2 \cos \left( 2 \pi \omega n \right)$, 
where $2t=t_1+t_2$,  reflects the Hofstadter butterfly pattern. The
lighter regions indicate the places where the bulk gap closes and the Lyapunov exponent vanishes i.e. the 
topologically trivial regions. The Hofstadter butterfly pattern shows loci in parameter space where available 
energy states exist in the limit $\delta=t_1-t_2=0$. These loci seed topological phases, which 
occupy larger regions of the phase diagram as $\delta$ grows larger between panels (a) $\delta=0.1$ and (b) $\delta=0.2$.}
\label{fig:butterfly}
\end{center}
\end{figure}

One-dimensional fermionic systems are especially sensitive to non-uniform potentials and disorder. The
similarity transformation quantifies the fact that the tendency to localize fermions directly competes
with the topological phases. The transfer matrix formalism is particularly well-suited to the study of
spatially varying potentials. The topological phases can be determined by considering the
eigenvalues of $\prod_i \mathcal{A}_i$, where the product runs over a full unit cell of the system in question.

{\bf Check line below whether it's right or left.}
More precisely, if both eigenvalues of $\mathcal{A}$ have a magnitude less than 1, this
indicates that the $a_i/A_i$ zero mode decays as it goes into the bulk from the right.
Since the transfer matrix governing the $b_i/B_i$ mode is given by $\mathcal{A}^{-1}$, in
this case the $b_i/B_i$ mode would be localized to the right-hand side of the system. If
however $\mathcal{A}$ has one eigenvalue that is larger than 1 in magnitude and another
that is smaller than 1, then the zero modes are not normalizable and thus this case corresponds to the trivial phase.

Physically, the topology of the Kitaev-like ladder is dictated by the competition
between the coupling parameters $\mu/t$ and $\delta = \Delta / t$. The
parameter $\delta$ breaks the symmetry between the $a_1$ and $B_1$ modes
at the left end of the chain. In the limit $\delta = 0$, the Hamiltonian treats the
$a_i / A_i$ and $b_i / B_i$ modes symmetrically, and thus must be trivial since the
topologically non-trivial modes are characterized by a \emph{single} end mode. For
$0 < \delta \leq 1$, the coupling of the $a_1$ mode to the rest of the chain is
weakened relative to the $B_1$ mode, and thus in this case any topological phase will have $N_S = 1$.

We begin our investigation of inhomogenously coupled SSH chains with a particulary simple case which illustrates the relevant physics. Consider the
case in which $\delta = -1$ (recall that we have set $t=1$) and $\mu_n$ is some arbitrary configuration (either periodic,
quasiperiodic, or disordered). As discussed above, this case can be considered as a single connected chain as in Fig.~\ref{fig:ladder} (d).
In this case, the product of transfer matrices can be carried out explicity
\begin{equation}
\prod_{n=1}^N \mathcal{A}_{n} = \frac{1}{\left( 2 t \right)^N} \prod_{n=1}^N \mu_i \left(
  \begin{array}{cc}
    1 & 0 \\
    2t/\mu_1 & 0 \\
  \end{array}
\right).
\end{equation}
The eigenvalues of the resultant matrix are $\frac{1}{\left( 2 t \right)^N} \prod_{n=1}^N \mu_n$ and zero.
The phase is topological provided that the nonzero eigenvalue has magnitude less than 1. The condition
that the system is topological can be expressed as
\begin{equation}
\langle \ln \left( | \mu_n |/t \right) \rangle \leq \ln 2,
\end{equation}
a condition that is applicable to periodic, quasiperiodic, and disordered potentials. We point out that
this special case ($\delta = -1$) can be mapped to the disordered quantum Ising model. In this
correspondence, the topological phase in the fermionic system corresponds to the ferromagnetic phase in the spin model.

In discussing modulations of the inter-chain coupling $\mu_n$, of particular note is the case in which
\begin{equation}
\mu_n/t = V/t + 2 \cos \left( 2 \pi \omega n \right).
\label{eq:quasi}
\end{equation}
This variation is of particular interest in the context of localization physics and arises in the
problem of an electron on a two-dimensional lattice in a magnetic field. The spectrum of the
latter is fractal and is known as Hofstadter's butterfly. The fractal pattern arises due to $\omega$ continuously 
tuning through quasi-periodic and periodic modulations with respect to the underlying lattice. In the present notation, Hofstadter's
butterfly corresponds to the regions in which the Lyapunov exponent for Eq. (\ref{eq:quasi}) and
$\delta$ are zero.  Given the similarity transformation discussed above, knowing Lyapunov exponents in the Hofstadter case 
enables us to derive Lyapunov exponents for  non-zero $\delta$. This in turn enables us to determine the regions of 
parameter space that support localized end bound states and are thus topological.

We have plotted the topological phase diagram in the $V-\omega$ plane for two values of
$\delta$ in Fig.~\ref{fig:butterfly}. As $\delta$ is increased from zero, the mapping to Hofstadter's model 
indicates that the regions corresponding to the lowest Lyapunov
exponent get filled in first. Thus, as $\delta$ is increased from zero, regions of the phase
diagram nearest to available energy states reflected in the Hofstadter butterfly pattern become topological first. As noted above,
increasing $\delta$ tends to isolate an $a_i/A_i$ mode, giving rise to more regions which are topological.
The phase diagram shown in Fig.~\ref{fig:butterfly} ultimately arises as a result of the competition
between the localizing potential $\mu_n$ given by Eq.~(\ref{eq:quasi}) and
the effects of nonzero $\delta$.

Our treatment above of spatial variations in the inter-chain coupling $\mu$ is generic and allows for different forms of variation.
Disorder in $\mu$ is another natural choice. The extensive treatments of such a case for the Kitaev chain~\citep{DeGottardi13}
immediately translate to the phase diagrams expected for the SSH ladder system.

\section{Finite Size}
\label{sec:finitesize}

In experimental set-ups, for instance cold-atomic optical lattices, the number of lattice sites in the
system is typically quite small due to limitations of experimental techniques. Consequently, finite
size effects are a significant consideration in experimental observation of topological phases. 
We thus derive the structure of edge mode wavefunctions for a system of size '$L$' and demonstrate
how a phase boundary can be effectively obtained through finite-size analyses.

For the reduced SSH ladder model -- the analogue of the Kitaev wire discussed in Sec.~\ref{sec:kitaev} -- it is sufficient to consider the Fourier space
Hamiltonian
\begin{eqnarray}\label{eq:H1rho}
 \hat{H}^{1}_k=\left(\begin{array}{cc}
0 & \rho(k)\\
\rho^{*}(k) & 0 \\
\end{array}\right)
\end{eqnarray}
 where $\rho(k)=t_1e^{ika}+t_2e^{-ika}+\mu$ and $\rho(k)= |\rho(k)|e^{i\phi(k)}$.
This Hamiltonian, Eq.~(\ref{eq:H1rho}), connects to the larger
Hamiltonian of Eq.~(\ref{eq:HamBlocks}) through a unitary transformation $U$ that renders
the latter off-diagonal and defines $A$ and $B$ sub-lattices that are linear
combinations of $a$ and $A$ ($b$ and $B$):
\begin{eqnarray}
U^{\dagger}\hat{H}_kU=\left(\begin{array}{cc}
  \hat{H}_k^{1} & 0\\
  0 & \hat{H}_k^{1}-2\mu\sigma_x\\
  \end{array}\right)
\end{eqnarray}
and $(A^{\dagger}_k,B^{\dagger}_k)=\frac{1}{\sqrt{2}}(a^{\dagger}_k+A^{\dagger}_k,b^{\dagger}_k+B^{\dagger}_k)$.

The eigenstates of Eq.~(\ref{eq:H1rho}), corresponding to two eigenvalues $\pm E$, are of the form:
 \begin{eqnarray}\label{eq:H1vec}
  |u_{k},\pm\rangle = \frac{1}{\sqrt{2}}\left(\begin{array}{c}
e^{-i\phi(k)}\\
\pm 1 \\
\end{array}\right).
\end{eqnarray}
Here, the phase $\phi(k)$ is implicitly given by
\begin{eqnarray}
\cot\phi(k)=\frac{t_1+t_2}{t_1-t_2}\cot k+\frac{\mu}{t_1-t_2}\csc k.
\end{eqnarray}
 We construct the edge mode states as linear combinations $|v_{k},\pm\rangle=C_{+}|u_{k},\pm\rangle + C_{-}|u_{-k},\pm\rangle$.
 Further, in terms of lattice wavefunctions~\citep{Delplace11}:
 \begin{eqnarray}\label{eq:uk}
&&  |u_{k},\pm\rangle = \sqrt{\frac{1}{2L}}\sum_{j=1}^{L}\left(\begin{array}{c}
e^{-i\phi(k)}\\
\pm 1 \\
\end{array}\right)\nonumber\\
&& \times (e^{ikj}|j,A\rangle, e^{ik(j+1)} |j+1,B\rangle)
\end{eqnarray}
where $A$ and $B$ denote the same sublattices as above.

To analyze the effects of the ladder's finite size we impose open boundary conditions and require that the wave functions vanish for sites $j=0, L+1$.
More precisely, we set
\begin{eqnarray}\label{eq:BCs}
  && \langle 0, B| v_{k},\pm\rangle=0\nonumber\\
  && \langle (L+1), A| v_{k},\pm\rangle =0.
\end{eqnarray}
 Enforcing these conditions leads us to identify phase boundaries as a function of $\delta, t, \mu$
 and the system size $L$ as
\begin{eqnarray}\label{eq:finalcond}
  \frac{t_1-t_2}{t_1+t_2-\mu}=\frac{\delta}{2t-\mu}= L+1.
\end{eqnarray}

As shown in Fig.~\ref{fig:Model12} (b), for the Kitaev wire analogue SSH ladder having finitely many lattice sites, the
slope of phase boundaries is linear in system size.
For large system sizes ($L\to\infty$), the boundaries become vertical and match the phase boundaries in the thermodynamic limit.
Further, enforcing Eq.~(\ref{eq:BCs}) enables us to obtain the form of the wavefunctions of zero energy edge modes:
\begin{eqnarray}\label{eq:FinEdge}
 && |v_{k_{\lambda}},\pm\rangle = \frac{1}{\sqrt{L}}\sum^{L}_{j=1}
  \left(\begin{array}{c}
\sinh(\lambda_{\xi} a(L+1+j))\\
\pm \sinh(\lambda_{\xi}(j+1)a) \\
\end{array}\right)\nonumber\\
&& \times (|j, A\rangle, |j+1,B\rangle )
\end{eqnarray}
where $\lambda_{\xi}$ is the inverse of the localization length $\xi$ for the edge state.
In other words, $\lambda_{\xi}$ is equivalent to the Lyapunov exponent $\gamma$
of Sec.~\ref{sec:kitaev}.

When $L>>\xi$ i.e. the system size is much larger than the localization length,  the difference in energy of the modes corresponding to the two
states in Eq.~(\ref{eq:FinEdge}) decreases exponentially fast with increasing $L$
so that for a large yet finitely-sized system the edge modes are
effectively degenerate states with vanishing energies. The 'bending' of the phase boundaries compared to the phase diagram of Fig.~\ref{fig:Model12} (a)
for a ladder with a finite size and the qualitative shape of the edge-mode wavefunction in Eq.~(\ref{eq:FinEdge}) are
germane to experimental studies of the SSH ladder system.

Further, we use the same formalism to compare the general SSH ladder to its decoupled $\mu\to 0$ limit.
Considering the case in which the decoupled system consists of two topological chains i.e. two chains hosting a pair of zero
egde modes, these four modes can be described by the wavefunctions
\begin{eqnarray}\label{eq:4x4FinSize}
 && |v^{\rm{ab}}_{k_{\lambda}},\pm; v^{\rm{AB}}_{k_{\lambda}},\pm\rangle=
  \sum_{j=1}^{L}\frac{(-1)^{j+1}}{\sqrt{L}}\left(\begin{array}{c}
  \sinh(\lambda_{\xi}(L+1-j))\\
  \pm \sinh(j\lambda_{\xi})\\
  \sinh(j\lambda_{\xi})\\
  \pm \sinh(\lambda_{\xi}(L-j+1))\\
  \end{array}\right)\nonumber\\
   && \times (|j,a\rangle, |j,b\rangle, |j,A\rangle, |j, B\rangle)
\end{eqnarray}
with $\lambda_{\xi}$ as above. An addition of a small inter-chain coupling $\mu$
can then be treated perturbatively. Consequently, we see that regardless
of the size of the SSH ladder, the four zero energy edge
states acquire an energy shift linear in $\mu$. In other words, these modes hybridize and the energy
spectrum does not possess any zero modes but rather two pairs of non-zero energy midgap states, as shown in Fig.~(\ref{fig:4To2}).
This observation is consistent with our previous discussion of pairs of zero
edge modes not being topologically protected, as in Sec.~\ref{sec:phases}.

\section{Outlook}
\label{sec:outlook}
In summary, we have analyzed the SSH ladder as a natural extension of the well-studied SSH chain,
paying particular regard to topological and edge state properties. We have charted the phase
diagram exhibited by such a ladder system and pinpointed the nature of the topological phases. Under
a restricted set of couplings, the ladder serves as an analog for the Kitaev chain and associated Majorana
physics. In this regime, we have investigated the effect of inhomogeneity and the possibility of a 'topological
Hofstadter butterfly phase diagram' for quasiperiodic variations. With an eye towards realizing these features
in a variety of experimental systems, we have discussed the role played by finite size effects.

One particularly elegant realization of localized edge modes in the context of cold atomic experiments
has already been achieved in Ref.~\citep{Meier16}, where these topological features of a single SSH chain
have been confirmed through direct imaging. The procedure
of directly detecting edge-localization entails loading condensate atoms on a particular lattice `site',
which in this experimental set-up corresponds to a discrete momentum state,  and suddenly turning on, or quenching,
the desired coupling between sites.  Observing consequent population decay, or the lack thereof, to the
neighboring `sites' through time-of-flight absorption imaging then
indicates whether or not a zero energy edge mode is supported for this point in the space of couplings. Such images
comprise a direct observation of an edge mode and confirm that the system is in a topologically
nontrivial phase. An extension of such methods to our SSH ladder would required a set-up capable of realizing
two coupled chains and the ability to tune through parameter space in order to access our predicted phase diagrams.

A similar procedure for observing edge modes has been used in a photonic system~\citep{Kitagawa12}, where a photon is initialized next to the
boundary between two topologically distinct quantum walks and observed to not spread ballistically when a bound edge state is present.
In metamaterials of Ref.~\cite{Peterson17}, the zero modes are found to localize
in the corners of the system and to be robust to mechanical deformations corresponding
to changes in the space of inter- and intra-chain couplings. These systems thus offer various means of manifesting Kitaev chain analogs through SSH ladder realizations.
The Majorana fermion bound states in these cases would translate to Dirac fermionic, bosonic, and
even classical mechanical localized modes confined to ends of the ladder.

Turning to inhomogeneity, the appearance of the Hofstadter butterfly is one of many exotic features arising from the rich mathematical structure of
the Harper equation and related quasi-periodicity.  Quantities of interest such as the wavefunctions and the density of states are known to have
multifractality~\citep{Hiramoto92}. Recently there have been works on realising a quasiperiodic system through
a Hofstadter Hamiltonian in cold atomic systems~\citep{Aidelsburger13, Miyake13}. Even though cold atomic systems are
well suited for realizations of complex structures such as the Hofstadter butterfly in contrast to
electronic systems, recent experimental studies centered on the Hofstadter Hamiltonian have not included any direct measurements of the fractal character of the
system's wavefunctions. Our model of coupled SSH chains provides a novel possibility of realizing this
striking self-similar diagram through observation of topological phases driven by direct time-of-flight imaging and related experimental techniques described above.

\begin{acknowledgements}
We are grateful to Diptiman Sen, Bryce Gadway, Alex An, Eric Meier and Taylor Hughes for useful discussions. This work is
supported by the National Science Foundation
under the grant DMR 0644022-CAR.
\end{acknowledgements}

\bibliography{SSHladders}

\end{document}